\documentclass[11pt]{amsart}
 
\usepackage{amscd,amssymb}
\usepackage{amsmath}
\usepackage{amsthm}
 
\usepackage[scaled]{helvet} % ss
\usepackage[pdftex]{graphicx,color}

\usepackage[cmtip,matrix,arrow]{xy}
\numberwithin{equation}{section}

\newtheorem {thm} 			{Theorem}     %this group shares the same counter

\newtheorem {Lemma}     [equation]     	{Lemma}

\newtheorem {Conjecture}[equation]	{Conjecture}
\newtheorem {prop}      [equation]      {Proposition}

\theoremstyle{definition}

\newtheorem {rem}       [equation]	{Remark}

\newcommand{\pr} {\smallskip\noindent{\bf Proof\,\,}}

\DeclareMathOperator{\Z}{\mathbb Z}

\newcommand{\R}{\mathbb {R}}

\newcommand{\fg}{\mathfrak{g}}

\newcommand{\cA} {\mathcal{A}}\newcommand{\cG} {\mathcal{G}}
\newcommand{\cD} {\mathcal{D}}

\begin{document}

\title[Fermionization and Convergence in Chern Simons Gauge Theory]{Fermionization and Convergent Perturbation Expansions in Chern-Simons Gauge Theory}
%    author one information
\author{Jonathan Weitsman}
\address{Department of Mathematics, Northeastern University, Boston, MA 02115}
\curraddr{}
\email{j.weitsman@neu.edu}
\thanks{Supported in part by NSF grant DMS 04/05670}
\thanks\today


\date{}

\begin{abstract}
We show that Chern-Simons gauge theory with appropriate cutoffs is equivalent, term by term in perturbation theory, to a Fermionic theory with a nonlocal interaction term.  When an additional cutoff is placed on the Fermi fields, this Fermionic theory gives rise to a convergent perturbation expansion.  This leads us to conjecture that Chern-Simons gauge theory also gives rise to convergent perturbation expansions, which would give a mathematically 
well-defined construction of the theory.\end{abstract}

\maketitle
\section{Introduction}\label{sec:introduction}
Chern-Simons gauge theory was studied by Witten \cite{witten} as a geometric context for the Jones polynomial using formal path integrals as follows.  Let $M$ be a compact three-manifold, and let $G$ be a compact simple Lie group.  Choose an invariant inner product on $\fg = Lie(G).$  The space $\Omega^1(M,\fg)$ of $\fg$-valued one-forms on $M$ can be identified with the space $\cA(M)$ of connections on the trivialized principal $G$-bundle on $P = M\times G \to M.$  In these terms, the Chern-Simons invariant of a connection $A \in \cA(M)$ is given by

$$ CS(A) = \frac{1}{4 \pi} {\rm tr~}\int_M A \wedge dA + \frac{2}{3} A^3.$$

Given $\lambda \in \Z,$ the partition function of the Chern-Simons quantum field theory is given schematically by
\begin{equation}\label{cspf}
\int_{\cA(M)} \cD A e^{-i\lambda CS(A)},
\end{equation}
\noindent where integration on $\cA(M)$ is a formal---and mysterious---operation.

The integrand in (\ref{cspf}) is invariant under the group $\cG = {\rm Aut~}(P) = {\rm Map ~} (M,G)$ of automorphisms of the bundle $P.$  The gauge fixed action was studied by Axelrod and Singer \cite{as}.  Suppose there exists a flat connection $A_0 \in \cA(M)$ such that $H^*(\Omega^*(M,\fg), d_{A_0})$ vanishes.\footnote{Here $d_{A_0}$ denotes the de Rham operator in the twisted de Rham complex corresponding to the bundle $M \times \fg = ad(P)$ and the connection $A_0.$}  Choose a Riemannian metric on $M.$ Choose also an orthonormal basis $e_\alpha$ for $\fg,$ and denote by $f_{\alpha\beta\gamma}$ the corresponding structure constants.  The gauge-fixed action is a function of a connection $A \in {\rm ker~} d_{A_0}^*$ and of two Fermi fields $c \in \Omega^0(M,\fg)$ and $C \in {\rm ker~}(d_{A_0}^*) \subset \Omega^2(M,\fg).$  It is given by \cite{as}

\begin{equation}\label{asaction}
S(A,c,C) = \frac{1}{2\pi}  \int_M \sum_\alpha \frac{1}{2}( A_\alpha \wedge (d_{A_0} A)_\alpha) - C_\alpha\wedge (d_{A_0} c)_\alpha + \frac{1}{6} \sum_{\alpha,\beta,\gamma} f_{\alpha\beta\gamma} (A_\alpha \wedge A_\beta \wedge A_\gamma - 6 C_\alpha \wedge A_\beta \wedge c_\gamma),
\end{equation}

\noindent and the gauge fixed partition function is given by

\begin{equation}\label{aspf}
Z_{\lambda}(M) = \int \cD A \cD c \cD C e^{-i\lambda S(A,c,C)}.
\end{equation}

Now the formal path integral appearing in (\ref{aspf}) is not in any sense well-defined.  However, it does give rise to a perturbation series by a variant of the usual Feynman procedure.  Axelrod and Singer show that each of the terms in this series is finite--in other words that the usual divergences appearing in perturbative quantum field theory do not appear in this case.  They also show that appropriate combinations of the terms in the perturbation series give rise to topological invariants of the three-manifold $M.$  The methods of \cite{as} do not address convergence of the perturbation series, and hence their results do not give a mathematical definition of the path integral.  Indeed the general expectation in Bosonic quantum field theories is that the perturbation series has radius of convergence equal to zero.

However, in \cite{w1} we showed that a cut-off version of Yang-Mills theory in four dimensions is equivalent, term-by-term in perturbation theory, to a Fermionic theory with nonlocal interactions.  This Fermionic theory, when given a further cutoff, gives rise to a {\em convergent} perturbation series.  The purpose of the present paper is to show that the methods of \cite{w1} apply also to Chern-Simons theory.  That is, a cut-off version of the action (\ref{asaction}) is equivalent, term-by-term  in perturbation theory, to a theory where the connection $A$ is replaced by a bilinear in Fermion fields (there is obviously no need to Fermionize $c$ and $C$ since they are already Fermions); and a further momentum cutoff placed on the Fermion fields yields a convergent perturbation series.  Since the perturbation series of Chern-Simons gauge theory, unlike that of Yang-Mills theory, is finite, we conjecture that it, too, is convergent.  However, our estimates are not uniform in the cutoff and are therefore not able to address this problem.

\subsection{The results of Axelrod and Singer}  We first describe in some more detail the results of Axelrod and Singer \cite{as}; we refer the reader to \cite{as} for more information.  

Let $\Delta_{A_0} = d_{A_0}^*d_{A_0} + d_{A_0} d_{A_0}^*$ be the Laplacian on $\Omega^*(M,\fg),$ and let $L: \Omega^*(M,\fg) \to \Omega^*(M,\fg)$ be the operator defined by 
$$L = d_{A_0}^* (\Delta_{A_0})^{-1}.$$

Denote the component of $L\circ *$ (where $*$ denotes the Hodge star operator) acting on $p-$forms by $L_p.$

If we choose an orthonormal framing of the tangent bundle $TM,$ we may view the kernels of $L_0$ and $L_1$ as a smooth functions on $M \times M - \Delta$ with values in $(\R^3 \otimes \fg) \otimes (\R^3 \otimes \fg);$ here $\Delta \subset M \times M$ denotes the diagonal.  Denote these functions
by $L_0(x,y)$ and $L_1(x,y)$ for $x,y \in M.$ 

Let $\chi \in C^\infty(\R)$ satisfy

\begin{itemize}
\item $1 \geq \chi \geq 0.$
\item $\chi' \geq 0.$
\item $\chi(x) = 0 {\rm ~if ~} x \leq 1.$
\item $\chi(x) = 1 {\rm  ~if ~}  x \geq 2.$
\end{itemize}

For $\epsilon > 0$ let $\chi_\epsilon(x) := \chi(\epsilon x),$  and for all 
$x,y \in M\times M - \Delta,$ and $i = 0,1,$ let 
$$L_i^\epsilon(x,y) := L_i(x,y) \chi_\epsilon(d(x,y)),$$ where $d(x,y)$ denotes the distance between $x$ and $y$ given by the Riemannian metric on $M.$  Then the functions $L_i^\epsilon$ extend to smooth functions on $M \times M,$ which we continue to denote by $L_i^\epsilon.$

The cut-off perturbation series of the action (\ref{asaction}) is given by the formal power series 
\begin{equation}\label{ps}
Z_{sc}(A_0;\lambda)\sum_{n=0}^\infty \frac{1}{\lambda^{n}} \Xi_n(\epsilon)\end{equation}
\noindent where $Z_{sc}(A_0;\lambda)$ is the semi-classical approximation to the partition function, involving Chern-Simons and torsion invariants of $A_0$ (see \cite{witten}), and where

\begin{equation}\label{asseries}
\Xi_n(\epsilon):= \frac{{R_0}^{3n}}{(3n)!}\frac{{R_I}^{2n}}{(2n)!}|_{A = 0, c=0, C=0},
\end{equation}
\noindent and $R_0,R_I$ are defined as follows. In terms of formal even variables $A_\alpha^i(x)$ and formal odd variables $c_\alpha(x), C_\alpha^{i,j}(x),$
$x\in M,$ $i,j = 1, 2, 3,$ $i<j,$ and $\alpha = 1, \dots, {\rm dim~}\fg,$  the polynomial $R_I(A,c, C)$ is  given by

$$R_I(A,c,C) := \frac{-i}{2\pi}\sum_{i,j,k, \alpha, \beta,\gamma} \int_M dx~
\epsilon_{ijk}f_{\alpha\beta\gamma} \Bigl(
\frac{1}{6}A_\alpha^i(x)A_\beta^j(x)A_\gamma^k(x)-
A_\alpha^i(x) c_\beta(x) C_\gamma^{j,k}(x)\Bigr),$$
\noindent and the formal differential operator $R_0$ is given by 

$$R_0 := {-2 \pi i}\sum_{i,j, \alpha, \beta} \int_{M \times M} dx dy 
~\Bigl ((L^\epsilon_1(x,y))_{i,j; \alpha, \beta} \frac{\delta}{\delta A_\alpha^i(x)}\frac{\delta}{\delta A_\beta^j(y)} -2 (L^\epsilon_0(x,y))_{i,j; \alpha, \beta} 
\frac{\delta}{\delta c _\alpha(x)}\frac{\delta}{\delta C_\beta^{i,j}(y)}\Bigr );$$

\noindent here we have used the notation $(L^\epsilon_k(x,y))_{i,j; \alpha, \beta}, k=0,1$ for the matrix elements of $L^\epsilon_k(x,y)$ in the basis given by the framing of the tangent bundle and the chosen basis $e_\alpha$ of $\fg.$

Then the key result of Axelrod and Singer is the following

\begin{thm}[Axelrod and Singer \cite{as} 1995]\label{asthm}
The limit 
$$
\Xi_n = \lim_{\epsilon\to 0} \Xi_n(\epsilon)$$

\noindent is finite for every $n.$
\end{thm}

Axelrod and Singer then show that the quantities $\Xi_n$ are topological invariants of $M.$  These ``finite-type'' invariants have been the focus of intensive research since the publication of   \cite{as}.
\subsection{Fermionization}

We now Fermionize Chern-Simons gauge theory, by a method similar to the one used in 
\cite{w1} in the case of Yang-Mills theory.  Morally, we replace the connection $A_\alpha^i$ with a bilinear in Fermi fields.  Let $H_i(x),$ $i=1,2,3,$ and
$\Psi_\alpha(x),$ $\alpha = 1,\dots, {\rm dim~} \fg,$ be complex Fermi fields.
The Fermi action is given by 

\begin{multline}\label{fermiac}
S_F^\epsilon(H_i,\Psi_\alpha,c_\alpha,C_\alpha^{i,j},
\bar{H}_i,\bar{\Psi}_\alpha,\bar{c}_\alpha,\bar{C}_\alpha^{i,j}
) =\\
 S_{F,0}(H_i,\Psi_\alpha,c_\alpha,C_\alpha^{i,j},\bar{H}_i,\bar{\Psi}_\alpha,\bar{c}_\alpha,\bar{C}_\alpha^{i,j})
+S_{F,I}^\epsilon(H_i,\Psi_\alpha,c_\alpha,C_\alpha^{i,j},\bar{H}_i,\bar{\Psi}_\alpha,\bar{c}_\alpha,\bar{C}_\alpha^{i,j})
\end{multline}

\noindent where

\begin{equation}\label{freefermiac}
S_{F,0}(H_i,\Psi_\alpha,c_\alpha,C_\alpha^{i,j},\bar{H}_i,\bar{\Psi}_\alpha,\bar{c}_\alpha,\bar{C}_\alpha^{i,j}) = \int_M dx\Bigl( 
\sum_i|H_i(x)|^2 + \sum_\alpha |\Psi_\alpha(x)|^2 + \sum_\alpha |c_\alpha(x)|^2
+ \sum_{\alpha,i,j} |C_\alpha^{i,j}(x)|^2\Bigr),
\end{equation}

\noindent and 
\begin{multline*}
S_{F,I}^\epsilon(H_i,\Psi_\alpha,c_\alpha,C_\alpha^{i,j},\bar{H}_i,\bar{\Psi}_\alpha,\bar{c}_\alpha,\bar{C}_\alpha^{i,j})=\\
\frac{i}{2\pi\sqrt{\lambda}} \sum_{i,j,k,\alpha,\beta,\gamma} 
\int_M dx ~\epsilon_{ijk}f_{\alpha\beta\gamma} 
\Bigl(
\frac{1}{(3!)^2} \bar{H}_i(x) \bar{\Psi}_\alpha(x)\bar{H}_j(x) \bar{\Psi}_\beta(x)\bar{H}_k(x) \bar{\Psi}_\gamma(x)
-\bar{H}_i(x) \bar{\Psi}_\alpha(x) 
\bar{c}_\beta(x)
\bar{C}_\gamma^{j,k}(x) 
\Bigr)\\
-
{2\pi i} \sum_{i,j,\alpha,\beta} \int_{M\times M} dx dy~
\Bigl(
(L_1^\epsilon(x,y))_{i,j;\alpha,\beta} H_i(x) \Psi_\alpha(x) H_j(y) \Psi_\beta(y)
-2
(L_0^\epsilon(x,y))_{i,j;\alpha,\beta} c_\alpha(x) C_\beta^{i,j}(y)
\Bigr).
\end{multline*}

To make further progress, we impose a cutoff on the Fermi fields, as in \cite{w1}.  It is convenient to do this by convolutions with approximate delta functions and step functions, as follows.

Let $\zeta \in C^\infty(\R)$ be an even function satisfying

\begin{itemize}
\item $\zeta \geq 0.$
\item $\int_{0}^\infty x^2 \zeta(x) dx  = \frac{1}{4 \pi}.$
\item $\zeta'(x) \leq 0$ for $x > 0.$
\item $\zeta(x) = 0$ for $x \geq 1.$
\end{itemize}

Given $h > 0,$ define $\delta_h: M \times M \to \R$ by

$$\delta_h(x,y) := (2/h)^3\zeta(2d(x,y)/h).$$

Similarly let $Z \in C^\infty(\R)$ satisfy

\begin{itemize}
\item $Z \geq 0.$
\item $Z(x) = 1$ if $x \in [-1,1]$
\item $Z'(x) \leq 0$ if $x > 0.$
\item $Z(x) = 0$ for $|x| \geq 2.$
\end{itemize}

For all $h > 0,$ define $D_h: M \times M  \to \R$ by

$$D_h(x,y) := Z(d(x,y)/2h).$$

We now define the cut-off Fermi fields by

$$\Psi^h_\alpha(x) = \int_M \Psi_\alpha(y) \delta_h(y,x) dy,$$
$$H^h_i(x) = \int_M H_i(y) \delta_h(y,x) dy,$$
$$c^h_\alpha(x) = \int_M c_\alpha(y) \delta_h(y,x) dy,$$
$$C_\alpha^{i,j;h}(x) = \int_M C_\alpha^{i,j}(y) \delta_h(y,x) dy,$$
$$\bar{\Psi}^h_\alpha(x) = \int_M \bar{\Psi}_\alpha(y) \delta_h(y,x) dy,$$
$$\bar{H}^h_i(x) = \int_M \bar{H}_i(y) D_h(y,x) dy,$$
$$\bar{c}^h_\alpha(x) = \int_M \bar{c}_\alpha(y) \delta_h(y,x) dy,$$
$$\bar{C}_\alpha^{i,j;h}(x) = \int_M \bar{C}_\alpha^{i,j}(y) \delta_h(y,x) dy,$$

\noindent and define the cut-off Fermi action by

\begin{multline}\label{cutofffermiac}
S_F^{\epsilon,h}(H_i,\Psi_\alpha,c_\alpha,C_\alpha^{i,j},
\bar{H}_i,\bar{\Psi}_\alpha,\bar{c}_\alpha,\bar{C}_\alpha^{i,j}
) =\\
 S_{F,0}(H_i,\Psi_\alpha,c_\alpha,C_\alpha^{i,j},
\bar{H}_i,\bar{\Psi}_\alpha,\bar{c}_\alpha,\bar{C}_\alpha^{i,j}
)+
S_{F,I}^\epsilon(H_i^h,\Psi_\alpha^h,c_\alpha^h,C_\alpha^{i,j;h},
\bar{H}_i^h,\bar{\Psi}_\alpha^h,\bar{c}_\alpha^h,\bar{C}_\alpha^{i,j;h}
)
\end{multline}

Then the analog of the Fermionization theorem of \cite{w1} is the following

\begin{thm}\label{fermionization}

Each term of the perturbation series of the Fermi action $S_F^{\epsilon,h}$ coincides in the limit $h\to 0$ with the corresponding term $\Xi_n(\epsilon)$ of the 
perturbation series (\ref{ps}) of the cut-off, gauge-fixed Chern-Simons gauge theory.\end{thm}

\begin{rem}  As in \cite{w1}, it is possible to write down a Fermi theory which gives rise to a perturbation series identical with that of (\ref{ps}).  However this theory does not arise from a Lagrangian; the free correlation functions for the Fermi fields $H_i$ in this theory are given by 

$$ < \bar{H}_i(x) H_j(y) > = \delta_{ij} {\rm ~if~} x = y;$$
$$ < \bar{H}_i(x) H_j(y) > = 0  {\rm ~if~} x \neq y,$$

\noindent as might be expected from the limiting behavior of the correlations of the cut-off fields $\bar{H}_{i}^h, H_{i}^h.$\end{rem}

\subsection{Convergent perturbation theory}

As in other examples of purely Fermionic theories \cite{fmrs,gk,w1}, we expect the cut-off Fermionic action
(\ref{cutofffermiac}) to give rise to a convergent
perturbation series.
The analog of Theorem 2 of \cite{w1} is the following.

\begin{thm}\label{convergence}
The perturbation series corresponding to the action $S_F^{\epsilon,h}$
converges for all $\lambda \neq 0.$
\end{thm}

\begin{rem}
In fact it is not necessary to place an additional cut-off on the fields $c$ and $C$; the action 
$$
S_{F,0}(H_i,\Psi_\alpha,c_\alpha,C_\alpha^{i,j},
\bar{H}_i,\bar{\Psi}_\alpha,\bar{c}_\alpha,\bar{C}_\alpha^{i,j}
)+
S_{F,I}^\epsilon(H_i^h,\Psi_\alpha^h,c_\alpha,C_\alpha^{i,j},
\bar{H}_i^h,\bar{\Psi}_\alpha^h,\bar{c}_\alpha,\bar{C}_\alpha^{i,j}
)$$
\noindent also gives rise to a convergent perturbation series.\end{rem}

The convergence estimates used to prove Theorem \ref{convergence} are not uniform in $\epsilon$ and $h,$\footnote{A slight variation of our methods allows one to take the limit $\epsilon \to 0$ as long as $h$ remains finite; however, the theory remains a theory with cut-offs.} as is indeed the case for the Gross-Neveu model
\cite{fmrs,gk} and for Yang-Mills theory \cite{w1}.  In those asymptotically free theories, the coupling constant is adjusted to approach zero as the ultraviolet cutoff is removed, which in our case would correspond to the limit where $\epsilon$ and $h$ approach zero.  It is this fact which makes non-uniform estimates useful for the construction of the theory in \cite{fmrs,gk}.  The situation is different for Chern-Simons gauge theory, which is finite to all orders in perturbation theory.  So we make the following conjecture.

\begin{Conjecture}\label{convconj}
~\\ (a)  The perturbation series 

\begin{equation}\label{ps1}
\sum_{n=0}^\infty \frac{1}{\lambda^{n}} \Xi_n(\epsilon)\end{equation}

\noindent is convergent for all $\epsilon$ and all $\lambda \neq 0.$ \\

(b)  The limit series

\begin{equation}\label{ps2}
\sum_{n=0}^\infty \frac{1}{\lambda^{n}} \Xi_n\end{equation}

\noindent converges for all $\lambda \neq 0.$
\end{Conjecture}

Evidently a proof of Conjecture \ref{convconj} would depend on finer determinant estimates than the ones used by \cite{fmrs,gk,w1} and applied to our case in the proof of Theorem \ref{convergence}.
\begin{rem}  There is strong evidence from topology that Conjecture \ref{convconj} is true.  This is because the Chern-Simons-Witten invariants of three-manifolds can be written in terms of invariants of finite type, which morally correspond to combinations of connected diagrams in the Axelrod-Singer expansion.  For a fixed three-manifold $M,$ all but finitely many of these invariants vanish, so that we might expect there to be only a finite number of nonvanishing connected diagrams in the Axelrod-Singer expansion; thus the sum over all diagrams should be convergent.  A similar result holds for expectations of {\em gauge invariant} observables given by nonintersecting Wilson loops in $M=S^3,$ where the resulting invariant, which is the Jones polynomial, is given as a sum of finite-type Vassiliev invariants, all but finitely many of which vanish for a given link (See for example \cite{dbn}.) Fermionization gives a potential quantum field theoretic context for the finite-type property of these invariants.
\end{rem}
\section{Proof of Theorem \ref{fermionization}}

Recall that the terms $\Xi_n(\epsilon)$ of the perturbation series of Chern-Simons gauge theory with cutoff are given by

\begin{equation}
\Xi_n(\epsilon):= \frac{{R_0}^{3n}}{(3n)!}\frac{{(R_I)}^{2n}}{(2n)!}|_{A = 0, c=0, C=0},
\end{equation}
\noindent where $R_0$ and $R_I$ are given in terms of formal even variables $A_\alpha^i(x)$ and formal odd variables $c_\alpha(x), C_\alpha^{i,j}(x)$ by

$$R_I^\epsilon(A,c,C) := \frac{-i}{2\pi}\sum_{i,j,k, \alpha, \beta,\gamma} \int_M dx~
\epsilon_{ijk}f_{\alpha\beta\gamma} \Bigl(
\frac{1}{6}A_\alpha^i(x)A_\beta^j(x)A_\gamma^k(x)-
A_\alpha^i(x) c_\beta(x) C_\gamma^{j,k}(x)\Bigr)$$
\noindent and
$$R_0 := {-2 \pi i}\sum_{i,j, \alpha, \beta} \int_{M \times M} dx dy ~\Bigl ((L^\epsilon_1(x,y))_{i,j, \alpha, \beta} \frac{\delta}{\delta A_\alpha^i(x)}\frac{\delta}{\delta A_\beta^j(y)} -2 (L^\epsilon_0(x,y))_{i,j, \alpha, \beta} 
\frac{\delta}{\delta c _\alpha(x)}\frac{\delta}{\delta C_\beta^{i,j}(y)}\Bigr ).$$

Similarly the perturbation series of the Fermionic action $S_F^{\epsilon,h}$
is given by

$$\sum_{n=0}^\infty \frac{1}{\lambda^{n}} \Theta_n(\epsilon,h),$$

\noindent where
\begin{equation}\label{thetadef}
\Theta_n(\epsilon,h):= \int 
\cD H \cD\bar{H} \cD{\Psi} \cD \bar{\Psi} \cD c \cD\bar{c}\cD C \cD\bar{C} e^{S_{F,0}}
\frac{{(T_0^\epsilon)}^{3n}}{(3n)!}\frac{(T_I^h)^{2n}}{(2n)!},
\end{equation}

\noindent and where the polynomials $T_I^h$ and $T_0^\epsilon$ are given by

\begin{multline*}
T_I^h(H,\Psi,c,C) =\\
 \frac{i}{2 \pi} \sum_{i,j,k, \alpha, \beta,\gamma} \int_M dx~
\epsilon_{ijk}f_{\alpha\beta\gamma}
\Bigl(\frac{1}{(3!)^2} 
\bar{H}_i^h(x)\bar{\Psi}_\alpha^h(x)\bar{H}_j^h(x)\bar{\Psi}_\beta^h(x)\bar{H}_k^h(x)\bar{\Psi}_\gamma^h(x)-
\bar{H}_i^h(x) \bar{\Psi}^h_\alpha(x) \bar{c}^h_\beta(x) \bar{C}_\gamma^{j,k;h}(x)\Bigr)\end{multline*}

and 

$$T_0^\epsilon := -
{2\pi i} \sum_{i,j,,\alpha,\beta} \int_{M\times M} dx dy~
\Bigl(
(L_1^\epsilon(x,y))_{i,j;\alpha,\beta} H_i^h(x) \Psi^h_\alpha(x) H^h_j(y) \Psi_\beta^h(y)
-2
(L_0^\epsilon(x,y))_{i,j;\alpha,\beta} c_\alpha^h(x) C_\beta^{i,j;h}(y)
\Bigr).
$$

Thus 

\begin{multline}\label{thetaexp}
\Theta_n(\epsilon,h)= \frac{1}{(2n)!(3n)!}\int 
\cD H \cD\bar{H} \cD{\Psi} \cD \bar{\Psi} \cD c \cD\bar{c}\cD C \cD\bar{C}e^{S_{F,0}}\\
(-2 \pi i)^{3n} \sum_{k=1}^{6n} \sum_{i_k, j_k=1,2,3}
\sum_{\alpha_k,\beta_k=1,\dots {\rm dim~} \fg}
\int_{M^{6n}} dx_1 \dots dx_{6n} \int_{M^{6n}} dz_1\dots dz_{6n}\\
\Bigl(\prod_{l=1}^{3n} \Bigr[
(L_1^\epsilon(x_{2l},x_{2l-1}) )_{i_{2l},i_{2l-1}; \alpha_{2l},\alpha_{2l-1}}
H_{i_{2l}}^h(x_{2l})\Psi_{\alpha_{2l}}^h(x_{2l})
H_{i_{2l-1}}^h(x_{2l-1})\Psi_{\alpha_{2l-1}}^h(x_{2l-1})\\
-2
(L_0^\epsilon(x_{2l},x_{2l-1}) )_{i_{2l},i_{2l-1}; \alpha_{2l},\alpha_{2l-1}}
c_{\alpha_{2l}}^h(x_{2l}) C_{\alpha_{2l-1}}^{i_{2l},i_{2l-1};h}(x_{2l-1})
\Bigr ]\Bigr)\\
\Bigl(\frac{i}{2\pi}\Bigr)^{2n} 
\Bigl(\prod_{m=1}^{2n} \epsilon_{j_{3m}j_{3m-1}j_{3m-2}}
f_{\beta_{3m}\beta_{3m-1}\beta_{3m-2}} \delta(z_{3m},z_{3m-1})\delta(z_{3m},z_{3m-2})\\
\Bigl[
\frac{1}{(3!)^2}
\bar{H}_{j_{3m}}^h(z_{3m})
\bar{\Psi}_{\beta_{3m}}^h(z_{3m})
\bar{H}_{j_{3m-1}}^h(z_{3m-1})
\bar{\Psi}_{\beta_{3m-1}}^h(z_{3m-1})
\bar{H}_{j_{3m-2}}^h(z_{3m-2})
\bar{\Psi}_{\beta_{3m-2}}^h(z_{3m-2})\\-
\bar{H}_{j_{3m}}^h(z_{3m})
\bar{\Psi}_{\beta_{3m}}^h(z_{3m})
\bar{c}_{\beta_{3m-1}}^h(z_{3m-1})
\bar{C}_{\beta_{3m-2}}^{j_{3m-1},j_{3m-2};h}(z_{3m-2})
\Bigr]
\Bigr).\end{multline}

Expanding the products, we have 

\begin{multline*}
\Theta_n(\epsilon,h)= \frac{(-2\pi i)^{3n}}{(2n)!(3n)!}\Bigl(\frac{i}{2\pi}\Bigr)^{2n}\int 
\cD H \cD\bar{H} \cD{\Psi} \cD \bar{\Psi} \cD c \cD\bar{c}\cD C \cD\bar{C}e^{S_{F,0}}\\
\sum_{k=1}^{6n} \sum_{i_k, j_k=1,2,3}
\sum_{\alpha_k,\beta_k=1,\dots {\rm dim~} \fg}
\int_{M^{6n}} dx_1 \dots dx_{6n} \int_{M^{6n}} dz_1\dots dz_{6n}\\
\sum_{q=1}^{2n} {{2n} \choose q} {{3n}\choose{q+n}}
\prod_{l=1}^{q+n}
\Bigl[
(L_1^\epsilon(x_{2l},x_{2l-1}) )_{i_{2l},i_{2l-1}; \alpha_{2l},\alpha_{2l-1}}
H_{i_{2l}}^h(x_{2l})\Psi_{\alpha_{2l}}^h(x_{2l})
H_{i_{2l-1}}^h(x_{2l-1})\Psi_{\alpha_{2l-1}}^h(x_{2l-1})
\Bigr]\\
\prod_{l=q+n+1}^{3n}\Bigl[-2
(L_0^\epsilon(x_{2l},x_{2l-1}) )_{i_{2l},i_{2l-1}; \alpha_{2l},\alpha_{2l-1}}
c_{\alpha_{2l}}^h(x_{2l}) C_{\alpha_{2l-1}}^{i_{2l},i_{2l-1};h}(x_{2l-1})
\Bigr ]\\
\Bigl(\prod_{m=1}^{2n}
 \epsilon_{j_{3m}j_{3m-1}j_{3m-2}}
f_{\beta_{3m}\beta_{3m-1}\beta_{3m-2}} \delta(z_{3m},z_{3m-1})\delta(z_{3m},z_{3m-2})\Bigr)\\
\Bigl(\prod_{m=1}^q
\Bigl[
\frac{1}{(3!)^2}
\bar{H}_{j_{3m}}^h(z_{3m})
\bar{\Psi}_{\beta_{3m}}^h(z_{3m})
\bar{H}_{j_{3m-1}}^h(z_{3m-1})
\bar{\Psi}_{\beta_{3m-1}}^h(z_{3m-1})
\bar{H}_{j_{3m-2}}^h(z_{3m-2})
\bar{\Psi}_{\beta_{3m-2}}^h(z_{3m-2})\Bigr]
\Bigr)\\
\Bigl(\prod_{m=q+1}^{2n}
\Bigl[
-\bar{H}_{j_{3m}}^h(z_{3m})
\bar{\Psi}_{\beta_{3m}}^h(z_{3m})
\bar{c}_{\beta_{3m-1}}^h(z_{3m-1})
\bar{C}_{\beta_{3m-2}}^{j_{3m-1},j_{3m-2};h}(z_{3m-2})
\Bigr]
\Bigr).
\end{multline*}

Standard Feynman diagram techniques allow us to write $\Theta_n(\epsilon,h)$ in as a sum of terms corresponding to trivalent graphs.  When the ghost lines in such a graph are cut, we obtain a pair of graphs, one of which corresponds to the $\Psi$ and $\bar{\Psi}$ fields and one of which to the $H$ and $\bar{H}$ fields.  We use this fact to write

$$ \Theta_n(\epsilon,h) = \Theta_n^1(\epsilon,h) + \Theta_n^2 (\epsilon,h)$$

\noindent where $\Theta_n^1(\epsilon,h)$ is the sum of those terms in the
diagrammatic expansion of $\Theta_n(\epsilon,h)$ corresponding to pairs of
{\em identical} Feynman diagrams---that is, Feynman diagrams where the
combinatorics of the pairings of the $\Psi$ and $\bar{\Psi}$
fields are the same as those of the $H$ and $\bar{H}$ fields; we will prove that
in the limit $h \to 0$ the terms
appearing in $\Theta_n^1(\epsilon,h)$ approach the corresponding terms in
the diagrammatic expansion of $\Xi_n(\epsilon).$
The sum of the remaining terms in the diagrammatic expansion of
$\Theta_n(\epsilon,h),$ which we denote by $\Theta_n^2(\epsilon,h),$
consists of terms corresponding to pairs of Feynman diagrams where at least
one $H$ propagator is not matched by a corresponding $\Psi$ 
propagator.  We will see that in the limit $h \to 0,$ such an
``unmatched''  propagator will give rise to a factor of order
$O(h^3),$ so that $\lim_{h \to 0} \Theta_n^2(\epsilon,h) = 0.$
More precisely, we write

\begin{multline*} 
\Theta_n^1(\epsilon,h) =
\frac{(-2\pi i)^{3n}}{(2n)!(3n)!}\Bigl(\frac{i}{2\pi}\Bigr)^{2n}
\sum_{k=1}^{6n} \sum_{i_k, j_k=1,2,3}
\sum_{\alpha_k,\beta_k=1,\dots {\rm dim~} \fg}
\int_{M^{6n}} dx_1 \dots dx_{6n} \int_{M^{6n}} dz_1\dots dz_{6n}\\
\sum_{q=1}^{2n} {{2n} \choose q}{{3n}\choose{q+n}}
\sum_{\sigma \in S_{2q+2n}}
\prod_{l=1}^{q+n}
(L_1^\epsilon(x_{2l},x_{2l-1}) )_{i_{2l},i_{2l-1}; \alpha_{2l},\alpha_{2l-1}}
\prod_{l=q+n+1}^{3n}
(-2L_0^\epsilon(x_{2l},x_{2l-1}) )_{i_{2l},i_{2l-1}; \alpha_{2l},\alpha_{2l-1}}\\
\Bigl(\prod_{m=1}^{2n}
 \epsilon_{j_{3m}j_{3m-1}j_{3m-2}}
f_{\beta_{3m}\beta_{3m-1}\beta_{3m-2}} \delta(z_{3m},z_{3m-1})\delta(z_{3m},z_{3m-2})\Bigr)\\
\Bigl(\prod_{m=1}^q
\Bigl[
\frac{1}{(3!)^3}
\int \cD H \cD\bar{H} \cD{\Psi} \cD \bar{\Psi} e^{S_{F,0}}\\
H^h_{i_{\sigma(3m)}}(x_{\sigma(3m)})
\Psi^h_{\alpha_{\sigma(3m)}}(x_{\sigma(3m)})
H^h_{i_{\sigma(3m-1)}}(x_{\sigma(3m-1)})
\Psi^h_{\alpha_{\sigma(3m-1)}}(x_{\sigma(3m-1)})
H^h_{i_{\sigma(3m-2)}}(x_{\sigma(3m-2)})
\Psi^h_{\alpha_{\sigma(3m-2)}}(x_{\sigma(3m-2)})\\
\bar{H}_{j_{3m}}^h(z_{3m})
\bar{\Psi}_{\beta_{3m}}^h(z_{3m})
\bar{H}_{j_{3m-1}}^h(z_{3m-1})
\bar{\Psi}_{\beta_{3m-1}}^h(z_{3m-1})
\bar{H}_{j_{3m-2}}^h(z_{3m-2})
\bar{\Psi}_{\beta_{3m-2}}^h(z_{3m-2})\Bigr]
\Bigr)\\
\Bigl(\prod_{m=q+1}^{2n}
\Bigl[
\int \cD H \cD\bar{H} \cD{\Psi} \cD \bar{\Psi} e^{S_{F,0}}
H^h_{i_{\sigma(2q+m)}}(x_{\sigma(2q+m)})
\Psi^h_{\alpha_{\sigma(2q+m)}}(x_{\sigma(2q+m)})
\bar{H}_{j_{3m}}^h(z_{3m})
\bar{\Psi}_{\beta_{3m}}^h(z_{3m})
\Bigr]
\Bigr)\\
\int\cD c \cD\bar{c}\cD C \cD\bar{C}e^{S_{F,0}}
\Bigl(\prod_{l=q+n+1}^{3n}
{c}_{\alpha_{2l}}^h(x_{2l})
{C}_{\alpha_{2l-1}}^{i_{2l},i_{2l-1};h}(x_{2l-1})\Bigr)
\Bigl(\prod_{m=q+1}^{2n}\Bigl[
-\bar{c}_{\beta_{3m-1}}^h(z_{3m-1})
\bar{C}_{\beta_{3m-2}}^{j_{3m-1},j_{3m-2};h}(z_{3m-2})
\Bigr]
\Bigr).
\end{multline*}

To obtain an explicit expression for $\Theta_n^2(\epsilon,h),$ we proceed as
in \cite{w1}.  Given $\sigma, \tau \in S_{2q+2n},$ we say $\sigma \sim \tau$ if for every $m = 1, \dots, 3q,$ there exists $k \in 1,\dots, q$ such that 

\begin{equation}\label{sim1}
\sigma(m) \tau(m) \in \{3k,3k+1,3k+2\}
\end{equation}
\noindent and if for every $m = 3q+1, \dots, 2q+2n,$ we have
\begin{equation}\label{sim2}
\sigma(m) = \tau(m).
\end{equation}
If $\sigma \nsim \tau,$ there exists a smallest integer $m = m(\sigma,\tau)$ such that either (\ref{sim1}) or (\ref{sim2}) is false.  Then, as in \cite{w1},

\begin{multline}\label{theta2}
\Theta_n^2(\epsilon,h) =
\frac{(-2\pi i)^{3n}}{(2n)!(3n)!}\Bigl(\frac{i}{2\pi}\Bigr)^{2n}
\sum_{k=1}^{6n} \sum_{i_k, j_k=1,2,3}
\sum_{\alpha_k,\beta_k=1,\dots {\rm dim~} \fg}
\int_{M^{6n}} dx_1 \dots dx_{6n} \int_{M^{6n}} dz_1\dots dz_{6n}\\
\sum_{q=1}^{2n} {{2n} \choose q}{{3n}\choose{q+n}}
\Bigl( \frac{1}{3!}\Bigr)^{2q}
\sum_{\sigma,\tau \in S_{2q+2n}}
{\rm sgn}(\sigma,\tau)\\
\prod_{l=1}^{q + n}
(L_1^\epsilon(x_{2l},x_{2l-1}) )_{i_{2l},i_{2l-1}; \alpha_{2l},\alpha_{2l-1}}
\prod_{l=q+n+1}^{3n}
(-2L_0^\epsilon(x_{2l},x_{2l-1}) )_{i_{2l},i_{2l-1}; \alpha_{2l},\alpha_{2l-1}}\\
\Bigl(\prod_{m=1}^{2n}
 \epsilon_{j_{3m}j_{3m-1}j_{3m-2}}
f_{\beta_{3m}\beta_{3m-1}\beta_{3m-2}} \delta(z_{3m},z_{3m-1})\delta(z_{3m},z_{3m-2})\Bigr)
\prod_{l=1}^{2q+2n} \delta_{\alpha_{\sigma(l)},\beta_l}  \delta_{i_{\tau(l)},j_l}\\
\Bigl(
\prod_{\stackrel{l=1}{l\neq m(\sigma,\tau), \sigma^{-1}\circ \tau(m(\sigma,\tau))}}^{2q+2n}
\tilde{\delta}_h(x_{\sigma(l)},z_{f_{n,q}(l)})
\tilde{D}_h(x_{\tau(l)},z_{f_{n,q}(l)})\Bigr)\\
\tilde{\delta}_h (x_{\sigma(m(\sigma,\tau))}, z_{f_{n,q}(m(\sigma,\tau))})
\tilde{D}_h (x_{\tau\circ\sigma^{-1}\circ\tau (m(\sigma,\tau))},
                   z_{f_{n,q}(\sigma^{-1}\circ\tau (m(\sigma,\tau)))})\\
\tilde{D}_h (x_{\tau(m(\sigma,\tau))}, z_{f_{n,q}(m(\sigma,\tau))})
\tilde{\delta}_h (x_{\tau(m(\sigma,\tau))},
                        z_{f_{n,q}(\sigma^{-1}\circ\tau (m(\sigma,\tau)))})\\
\int\cD c \cD\bar{c}\cD C \cD\bar{C}e^{S_{F,0}}
\Bigl(\prod_{l=q+n+1}^{3n}
{c}_{\alpha_{2l}}^h(x_{2l})
{C}_{\alpha_{2l-1}}^{i_{2l},i_{2l-1};h}(x_{2l-1})\Bigr)
\Bigl(\prod_{m=q+1}^{2n}\Bigl[
-\bar{c}_{\beta_{3m-1}}^h(z_{3m-1})
\bar{C}_{\beta_{3m-2}}^{j_{3m-1},j_{3m-2};h}(z_{3m-2})
\Bigr]
\Bigr),
\end{multline}

\noindent where, as in \cite{w1}, ${\rm sgn}(\sigma,\tau)$ is a sign we will
not compute explicitly, where $f_{n,q}: \{1,\dots , 2q + 2n\}\to \Z$ is defined by 

\begin{equation}
f_{n,q}(l) = 
\begin{cases} l & \text{if $l \leq 3q$}\\
3q + 3 (l-3q) & \text{if $l > 3q,$}
\end{cases}
\end{equation}

\noindent and where we have written

$$\tilde{\delta}_h(x,y) := \int_M \delta_h(x,z) \delta_h(z,y) dz$$

\noindent and

$$\tilde{D}_h(x,y) := \int_M \delta_h(x,z) D_h(z,y) dz.$$

\begin{Lemma}\label{deltas}
The functions $\tilde{\delta}_h$ and $\tilde{D}_h$ are positive.  Furthermore, for $h$ sufficiently small,

\begin{itemize}
\item $\lim_{h\to 0} \tilde{\delta}_h = \delta$ (as elements of
${\mathcal D}^\prime(M)$).
\item $\tilde{\delta}_h \tilde{D}_h = \tilde{\delta}_h.$
\item  $||\tilde{D}_h(x,\cdot)||_{\infty} = 1$ for any $x \in M.$
\item  $||\tilde{\delta}_h(x,\cdot)||_{1} \leq C $ for any $x \in M,$  where $C$ is a constant independent of $h$ and $x.$
\item $||\tilde{\delta}_h\star\tilde{D}_h(x,\cdot)||_{1} = O(h^3)$ for any $x \in M,$
\end{itemize}

\noindent where $\tilde{\delta}_h\star\tilde{D}_h$ is the convolution

$$\tilde{\delta}_h\star\tilde{D}_h(x,y) :=
\int_M   \tilde{\delta}_h(x,z)\tilde{D}_h(z,y) dz.$$
\end{Lemma}

\begin{prop}  We have
$$\lim_{h \to 0} \Theta_n^2(\epsilon,h) = 0.$$
\end{prop}

{\bf Proof.}  By (\ref{theta2}), and using the fact that $M$ is compact and the $L_i^\epsilon$'s are bounded,

\begin{equation} 
|\Theta_n^2(\epsilon,h)| \leq C \sup_q
(\sup_{z\in M}||\tilde{\delta}_h(\cdot,z)||_1)^{6n-1}
(\sup_{z\in M}||\tilde{D}_h(\cdot,z)||_\infty)^{2q+2n-1} 
\sup_{z \in M}||\tilde{\delta}_h\star\tilde{D}_h(\cdot,z)||_{1}.\end{equation}

By Lemma \ref{deltas}, 

$$|\Theta_n^2(\epsilon,h)| = O(h^3).$$

It remains to show

\begin{prop}
$$\lim_{h \to 0} \Theta_n^1(\epsilon, h) = \Xi_n(\epsilon).$$
\end{prop}

{\bf Proof.}  We note that

$$\lim_{h\to 0} \int \cD H \cD\bar{H} \cD{\Psi} \cD \bar{\Psi} e^{S_{F,0}}
{H}_i^h(x) {\Psi}_\alpha^h(x) 
\bar{H}_j^h(y) \bar{\Psi}_\beta^h(y) = 
-\frac{\delta}{\delta A^i_\alpha(x)} A^j_\beta(y).$$

Hence
\begin{multline*}
\lim_{h\to 0} \Theta_n^1(\epsilon,h) =\\
\frac{(-2\pi i)^{3n}}{(2n)!(3n)!}\Bigl(\frac{-i}{2\pi}\Bigr)^{2n}
\sum_{k=1}^{6n} \sum_{i_k, j_k=1,2,3}
\sum_{\alpha_k,\beta_k=1,\dots {\rm dim~} \fg}
\int_{M^{6n}} dx_1 \dots dx_{6n} \int_{M^{6n}} dz_1\dots dz_{6n}\\
\sum_{q=1}^{2n} {{2n} \choose q}{{3n}\choose{q+n}}
\sum_{\sigma \in S_{2q+2n}}
\prod_{l=1}^{q+n}
(L_1^\epsilon(x_{2l},x_{2l-1}) )_{i_{2l},i_{2l-1}; \alpha_{2l},\alpha_{2l-1}}
\prod_{l=q+n+1}^{3n}
(-2L_0^\epsilon(x_{2l},x_{2l-1}) )_{i_{2l},i_{2l-1}; \alpha_{2l},\alpha_{2l-1}}\\
\Bigl(\prod_{m=1}^{2n}
 \epsilon_{j_{3m}j_{3m-1}j_{3m-2}}
f_{\beta_{3m}\beta_{3m-1}\beta_{3m-2}} \delta(z_{3m},z_{3m-1})\delta(z_{3m},z_{3m-2})\Bigr)\\
\Bigl(\prod_{m=1}^q
\Bigl[
\frac{1}{(3!)^2}
\frac{\delta}{\delta A^{i_{\sigma(3m)}}_{\alpha_{\sigma(3m)}}(x_{\sigma(3m)})}
\frac{\delta}{\delta A^{i_{\sigma(3m-1)}}_{\alpha_{\sigma(3m-1)}}(x_{\sigma(3m-1)})}
\frac{\delta}{\delta A^{i_{\sigma(3m-2)}}_{\alpha_{\sigma(3m-2)}}(x_{\sigma(3m-2)})}
A^{j_{3m}}_{\beta_{3m}}(z_{3m})
A^{j_{3m-1}}_{\beta_{3m-1}}(z_{3m-1})
A^{j_{3m-2}}_{\beta_{3m-2}}(z_{3m-2})
\Bigr]\Bigr)\\
\Bigl(\prod_{m=q+1}^{2n}
\Bigl[
\frac{\delta}{\delta A^{i_{\sigma(2q+m)}}_{\alpha_{\sigma(2q+m)}}(x_{\sigma(2q+m)})}
A^{j_{2q+m}}_{\beta_{2q+m}}(z_{3m})
\Bigr]
\Bigr)\\
\int\cD c \cD\bar{c}\cD C \cD\bar{C}e^{S_{F,0}}
\Bigl(\prod_{l=q+n+1}^{3n}
{c}_{\alpha_{2l}}(x_{2l})
{C}_{\alpha_{2l-1}}^{i_{2l},i_{2l-1}}(x_{2l-1})\Bigr)
\Bigl(\prod_{m=q+1}^{2n}\Bigl[
-\bar{c}_{\beta_{3m-1}}(z_{3m-1})
\bar{C}_{\beta_{3m-2}}^{j_{3m-1},j_{3m-2}}(z_{3m-2})
\Bigr]
\Bigr)\\=\Xi_n(\epsilon).
\end{multline*}

\section{Proof of Theorem \ref{convergence}}

Recall the explicit expression for the terms of the perturbation series of the action $S_F^{\epsilon,h}.$  This perturbation series is given by

$$\sum \frac{1}{\lambda^n}\Theta_n(\epsilon,h)$$

\noindent where (see (\ref{thetadef}) and (\ref{thetaexp}))

\begin{multline*} 
\Theta_n(\epsilon, h) =
\frac{1}{(2n)!(3n)!} \int \cD \Psi \cD H \cD c \cD C\cD\bar{\Psi} \cD\bar{H}
\cD\bar{c} \cD \bar{C} \exp(S_{F,0}) \\
\Bigl(
\frac{i}{2\pi\sqrt{\lambda}} \sum_{i,j,k,\alpha,\beta,\gamma} 
\int_M dx ~\epsilon_{ijk}f_{\alpha\beta\gamma}\Bigl[ 
\frac{1}{(3!)^2} \bar{H}_i^h(x) \bar{\Psi}^h_\alpha(x)\bar{H}^h_j(x) 
\bar{\Psi}^h_\beta(x)\bar{H}^h_k(x) \bar{\Psi}^h_\gamma(x)
-\bar{C}_\alpha^{i,j;h}(x) \bar{H}^h_k(x) \bar{\Psi}^h_\beta(x) \bar{c}^h_\gamma(x)\Bigr]
\Bigr)^{2n}\\
\Bigl(
-
{2\pi i} \sum_{i,j,,\alpha,\beta} \int_{M\times M} dx dy~\Bigl[
(L_1^\epsilon(x,y))_{i,j;\alpha,\beta} H^h_i(x) \Psi^h_\alpha(x) H^h_j(y) 
\Psi^h_\beta(y)
-2
(L_0^\epsilon(x,y))_{i,j;\alpha,\beta} c^h_\alpha(x) C_\beta^{i,j;h}(y)\Bigr]
\Bigr)^{3n}.
\end{multline*}
Theorem \ref{fermionization} follows from the following estimate.

\begin{prop}\label{estimate} There exists a constant $C=C(\epsilon,h)>0$
such that 

$$|\Theta_n(\epsilon, h)| \leq \frac{C^n}{(2n)!(3n)!}.$$
\end{prop}

To prove Proposition \ref{estimate}, we note that $\Theta_n(\epsilon,h)$
is a sum of $O(C^n)$ terms, each of which is (up to a constant of order $C^n$) of the form

\begin{multline}\label{term}
\frac{1}{(2n)!(3n)!}\int \cD \Psi \cD H \cD c \cD C\cD\bar{\Psi} \cD\bar{H}
\cD\bar{c} \cD \bar{C}\exp(S_{F,0}) \\
\int_{M^{6n}} dx_1 dy_1\dots dx_{3n} dy_{3n} \int_{M^{2n}} dz_1\dots dz_{2n}
\prod_{l=1}^{\frac{3q+p}{2}}
(L_1^\epsilon(x_l,y_l))_{i_l,j_l;\alpha_l,\beta_l}
{{H}_{i_l}^h(x_l)} 
{{\Psi}^h_{\alpha_l}(x_l)}
{{H}_{j_l}^h(y_l)} 
{{\Psi}_{\beta_l}^h(y_l)}\\
\prod_{l={\frac{3q+p}{2}+1}}^{3n}
(-2L_0^\epsilon(x_l,y_l))_{i_l,j_l;\alpha_l,\beta_l}
{{c}_{\alpha_l}^h(x_l)} 
{{C}_{\beta_l}^{i_l,j_l;h}(y_l)}
\\
\prod_{m=1}^{q} 
\bar{H}^h_{p_m}(z_m) 
\bar{\Psi}_{\theta_m}^h(z_m) 
\bar{H}_{q_m}^h(z_m) 
\bar{\Psi}_{\iota_m}^h({z_m}) 
\bar{H}_{r_m}^h(z_m)
\bar{\Psi}_{\sigma_m}^h(z_m)\\
\prod_{m=q+1}^{2n} 
\bar{H}_{p_m}^h(z_m) 
\bar{\Psi}_{\theta_m}^h(z_m) 
\bar{c}_{\iota_m}^h({z_m}) 
\bar{C}_{\sigma_m}^{q_m,r_m;h}(z_m),
\end{multline}

\noindent where $p+q=2n.$ 

The Berezin integral appearing in (\ref{term}) is 

\begin{multline*}
B:= 
\int \cD \Psi \cD H \cD c \cD C\cD\bar{\Psi} \cD\bar{H}
\cD\bar{c} \cD \bar{C}\exp(S_{F,0}) 
\int_{M^{6n}} dx_1 dy_1\dots dx_{3n} dy_{3n} \int_{M^{2n}} dz_1\dots dz_{2n}\\
\prod_{l=1}^{\frac{3q+p}{2}}
{{H}_{i_l}^h(x_l)} 
{{\Psi}^h_{\alpha_l}(x_l)}
{{H}_{j_l}^h(y_l)} 
{{\Psi}_{\beta_l}^h(y_l)}
\prod_{l={\frac{3q+p}{2}+1}}^{3n}
{{c}_{\alpha_l}^h(x_l)} 
{{C}_{\beta_l}^{i_l,j_l;h}(y_l)}
\\
\prod_{m=1}^{q} 
\bar{H}^h_{p_m}(z_m) 
\bar{\Psi}_{\theta_m}^h(z_m) 
\bar{H}_{q_m}^h(z_m) 
\bar{\Psi}_{\iota_m}^h({z_m}) 
\bar{H}_{r_m}^h(z_m)
\bar{\Psi}_{\sigma_m}^h(z_m)\\
\prod_{m=q+1}^{2n} 
\bar{H}_{p_m}^h(z_m) 
\bar{\Psi}_{\theta_m}^h(z_m) 
\bar{c}_{\iota_m}^h({z_m}) 
\bar{C}_{\sigma_m}^{q_m,r_m;h}(z_m).
\end{multline*}

This Berezin integral is the inner product of two elements of
$$\bigwedge^{6q+4p} \left(L_2(M) \otimes \left(\R^3 \oplus \fg \oplus \fg \oplus \left (\fg\otimes \R^3 \otimes \R^3\right)\right)\right    ),$$
\noindent and is bounded by 

$$|B| \leq \bigl({\rm sup}_{x \in M} ||\delta_h(x,\cdot)||_{L_2(M)}\bigr)^{9q+7p}
\bigl({\rm sup}_{x \in M} ||D_h(x,\cdot)||_{L_2(M)}\bigr)^{3q+p}.$$

Since the kernels $L_i^\epsilon$ are smooth, Proposition \ref{estimate} follows.

\section{Remarks}
\subsection{Correlation functions}  As in \cite{w1}, 
the generating function for correlation functions of the gauge fields, which
is obtained by adding a term of the form 
$\int_M \sum_{i,\alpha} J^i_\alpha(x) A^i_\alpha(x) dx$ to the Chern-Simons
Lagrangian, can be Fermionized by adding the term 
$\int_M \sum_{i,\alpha} J^i_\alpha(x) \bar{H}_i(x) \bar{\Psi}_\alpha(x) dx$ to the
Fermionized action.
\subsection{Yang-Mills and QCD in three dimensions}  Our techniques apply just as well to a Lagrangian obtained by adding a Yang-Mills term 

$$ S(A) = \frac{1}{\lambda^2}\int_{M}|F(A)|^2$$

\noindent to the Chern-Simons Lagrangian.  As in \cite{w1}, such a Lagrangian
is equivalent to the Lagrangian

$$S(A,F) = \frac{1}{\lambda^2}(||F||^2  + 2i \langle F, dA\rangle +  2i \langle F, [A, A]\rangle ) + i k CS(A)$$

\noindent where $F \in \Omega^2(M,\fg)$ is a conjugate field.  The theory then has a cubic interaction term and can be Fermionized by the same method we have used for pure Chern-Simons
gauge theory.  These ideas work also for pure Yang-Mills theory (with no Chern-Simons term) in three dimensions.  The addition of Fermionic matter fields can likewise be
accommodated by the same techniques.

Similar techniques should also apply to two-dimensional gauge theories.

\subsection{String field theory}  I believe that our techniques should also
give a Fermionization of Witten's string field theory.  Recall that the 
string field theory Lagrangian is given by

$$S_{\rm sft}(A) = \int (A * Q A + \frac23 A * A * A),$$
\noindent where $A=A(\varphi,b,c)$
is the string field, which is a function
of a bosonic field $\varphi$ and two ghosts $b$ and $c,$ and the operator $Q$ and the operations
$\int$ and $*$ are defined in \cite{wittensft}.  Imposing a gauge condition
reduces the quadratic part of $S_{\rm sft}$ (up to a constant) to a positive-definite form.  One
can then write, as in this paper and in \cite{w1}

$$A = H(\pi_+\varphi,\pi_+b,\pi_+c)\Psi(\pi_-\varphi,\pi_-b,\pi_-c)$$

\noindent where $H$ and $\Psi$ are fermionic fields, $\pi_+$ is the
operator on Fock space induced by the projection
$\pi_+: L_2([0,1]) \to L_2([0,\frac12]),$ and $\pi_-$ is the operator
induced on Fock space by the projection
$\pi_-: L_2([0,1]) \to L_2([\frac12,1]).$\footnote{The addition of a gauge group to the string field by factors attached to the string edges can be accommodated (in the case of $SU(n)$ or $SO(n)$) by taking the fermionic string fields $H$ and $\Psi$ to have values in the fundamental representation of the gauge group.}  There are
various technical problems associated with the ghost current anomaly,
but I hope that with a proper cutoff (such as that of \cite{w2}) 
this theory can also be shown to yield a convergent perturbation series.


\begin{thebibliography}{asdf}
\bibitem{as} S. Axelrod, I. M. Singer.  J. Differential Geom.  39  (1994),  no. 1, 173--213

-- Proceedings of the XXth International Conference on Differential Geometric Methods in Theoretical Physics, Vol. 1, 2 (New York, 1991),  3--45, World Sci. Publ., River Edge, NJ, 1992.

\bibitem{dbn} D. Bar Natan, Topology 34 (1995) 423-472

-- math.GT/0408182.

\bibitem{fmrs} J.Feldman, J. Magnen, V. Rivasseau, R. Seneor, {\em Commun. Math. Phys.} {\bf 103}, 67-105 (1986)
\bibitem{gk} K. Gawedzki, A. Kupiainen, {\em Commun. Math. Phys.} {\bf 102}, 1-30 (1985)
\bibitem{gj} J. Glimm, A. Jaffe.  Quantum Physics.  Springer, 1987

\bibitem{salmhofer} M. Salmhofer.  Renormalization.  Springer Verlag, 1999.
\bibitem{w1} J. Weitsman,  Fermionization, Convergent Perturbation Theory, and Correlations  in the Yang-Mills Quantum Field Theory in four dimensions.  Preprint arxiv:0902.0096
\bibitem{w2} J. Weitsman,  Measures on Banach Manifolds, Random Surfaces, and Nonperturbative String Field Theory with Cut-offs. arXiv:0807.2069 
\bibitem{witten} E. Witten, {\em Commun. Math. Phys.} {\bf 121},  359 (1988)
\bibitem{wittensft} E. Witten, {\em Nucl. Phys.} {\bf B268} 253 (1986)

\end{thebibliography}
\end{document}